\documentclass[manuscript]{aastex}

\usepackage[]{graphicx}
\usepackage{amsmath,amssymb}

\def\mf{\mathbf}

\newcommand{\mfbar}[1]{\mf{\bar{#1}}}
\newcommand{\mfhat}[1]{\mf{\hat{#1}}}
\newcommand{\bmu}{\boldsymbol{\mu}}
\newcommand{\blam}{\boldsymbol{\Lambda}}
\newcommand{\refeq}[1]{Equation  (\ref{#1})}

\newcommand{\leftexp}[2]{{\vphantom{#2}}^{#1}\!{#2}}

\begin{document}

\title{Sequential Covariance Calculation for Exoplanet Image Processing}

\author{Dmitry Savransky}
\affil{Sibley School of Mechanical and Aerospace Engineering, Cornell University, Ithaca, NY 14853, USA}

\email{ds264@cornell.edu}

\begin{abstract}
Direct imaging of exoplanets involves the extraction of very faint signals from highly noisy data sets, with noise that often exhibits significant spatial, spectral and temporal correlations.  As a results, a large number of post-processing algorithms have been developed in order to optimally decorrelate the signal from the noise.  In this paper, we explore four such closely related algorithms, all of which depend heavily on the calculation of covariances between large data sets of imaging data.  We discuss the similarities and differences between these methods, and demonstrate how the use sequential calculation techniques can significantly improve their computational efficiencies.
\end{abstract}


\section{Introduction}
In the last twenty years, the existence of exoplanets (planets orbiting stars other than our own sun) has gone from conjecture to established fact.  The accelerating rate of exoplanet discovery has generated a wealth of important new knowledge, and is due mainly to the development and maturation of a large number of technologies that drive a variety of planet detection methods.  We have now confirmed nearly one thousand exoplanets, with several thousand additional candidates already identified \citep{batalha2013planetary}.  The majority of these detections were made indirectly, by searching for the effects of a planet on its target star either via gravitational interaction as in Doppler spectroscopy surveys, or for direct blocking of starlight as in transit photometry.

While indirect detection methods have proven very successful in discovering exoplanets, they are dependent on collecting multiple orbits of data and are thus biased towards short-period planets.  Direct imaging, on the other hand, is biased towards planets on larger orbits, making it highly complementary to the indirect methods.  Together, these techniques can significantly advance our understanding of planet formation and evolution at all orbital scales.  Additionally, direct imaging provides the most straightforward way of getting planet spectra, which are invaluable to the study of planetary and atmospheric compositions and can serve as proxies for planet mass measurements \citep{barman2011clouds}.  For these reasons, there is now a concentrated effort to develop direct exoplanet imaging capability, both for ground based telescopes and for future space observatories.  At the same time, there are multiple groups working on the post-processing aspect of planetary imaging, and developing more advanced algorithms for extracting planetary signals from highly noisy backgrounds.

Giant planets are typically millions of times fainter than their host stars, with the very brightest (and youngest) emitting approximately $10^5$ times less light than their parent stars in the infra-red.  Earth-like planets reflect one part in 10 billion of light from their host stars.  Therefore, we require specially-designed, dedicated instrumentation to directly image planets.  This is usually some form of coronagraph and wavefront control system, with many different iterations currently under development.  While the instrument is designed to take care of both the diffraction and dynamic range problems that make exoplanet imaging so difficult, the final science images still contain some residual noise at approximately the level of the expected planet signal.  This noise comes from a variety of sources, including imperfections in the instrument optics, non-common path errors in instruments with dedicated wavefront sensors, and (especially for ground-based instruments) uncorrected residuals from an adaptive optics (AO) system working to counter the effects of atmospheric turbulence.  The different types of noise also have different spatial and temporal distributions - while detector readout noise and shot noise are completely random in space and time, noise from optical aberrations and AO residuals (referred to as speckles) will be correlated on the scale of the planet signal and will often persist through many subsequent images.

\begin{figure}
	\begin{center}
	\includegraphics[width=0.33\textwidth]{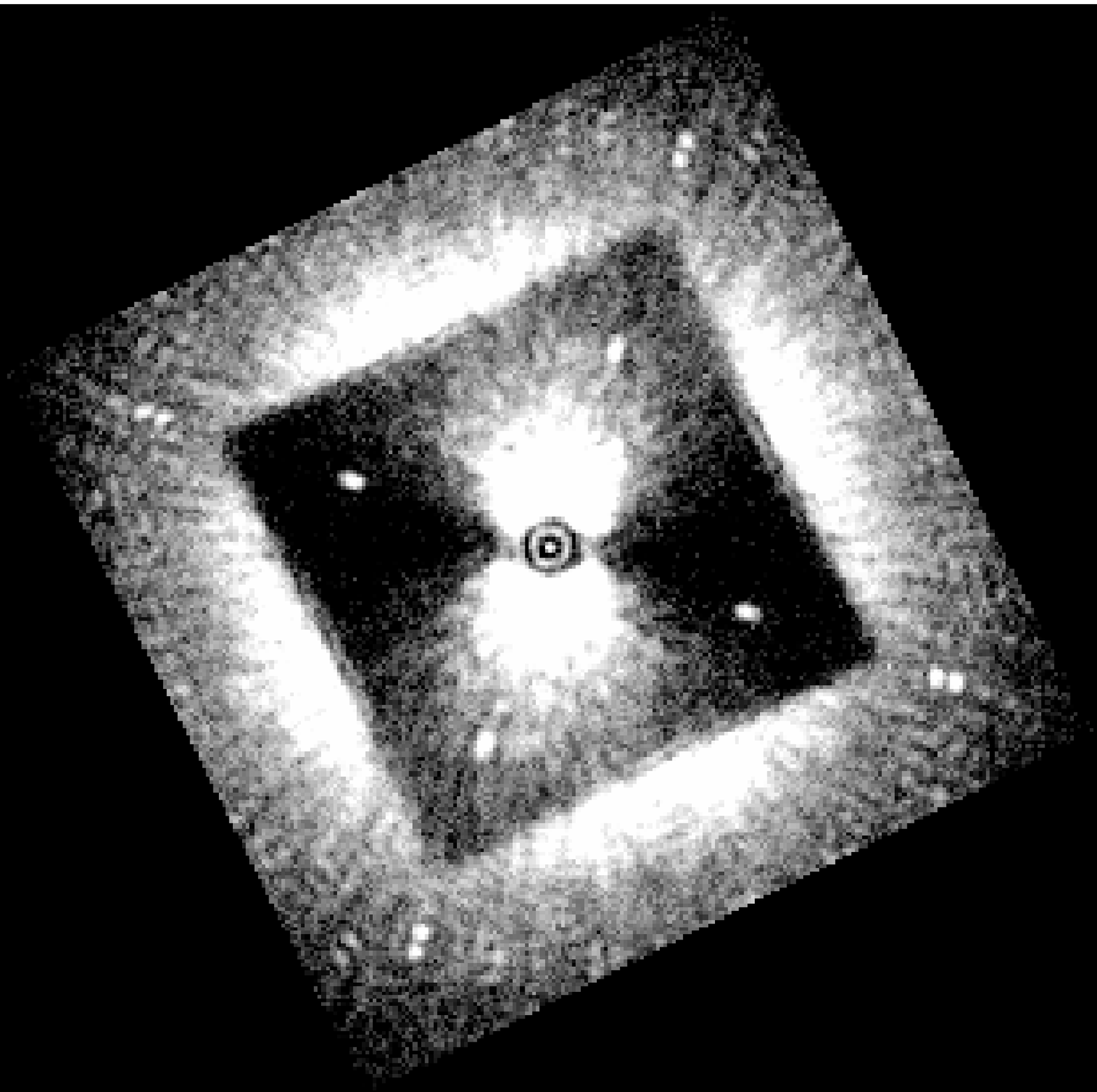}
	\hspace{-2ex}
	\includegraphics[width=0.33\textwidth]{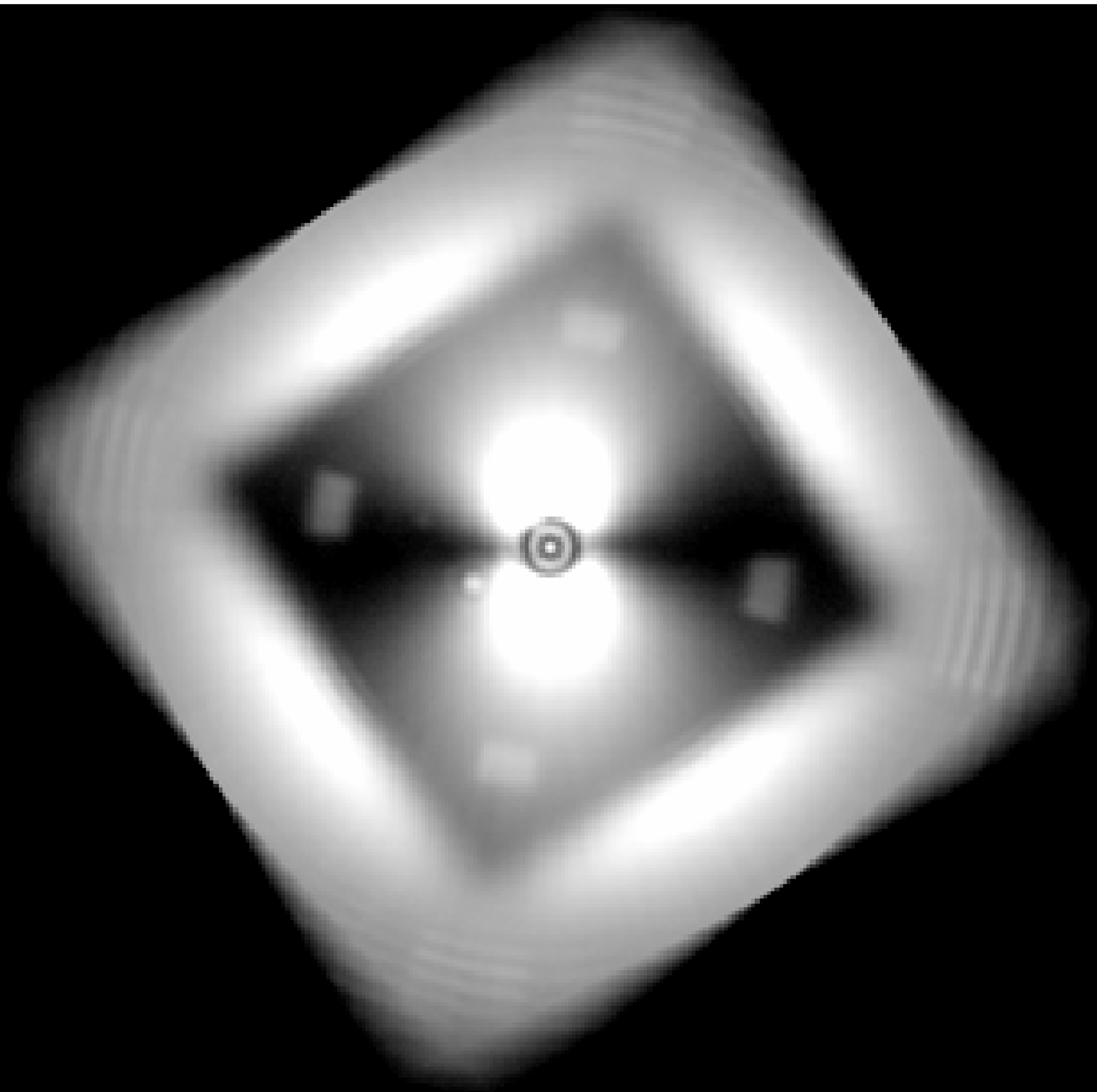}
	\hspace{-2ex}
	\includegraphics[width=0.33\textwidth]{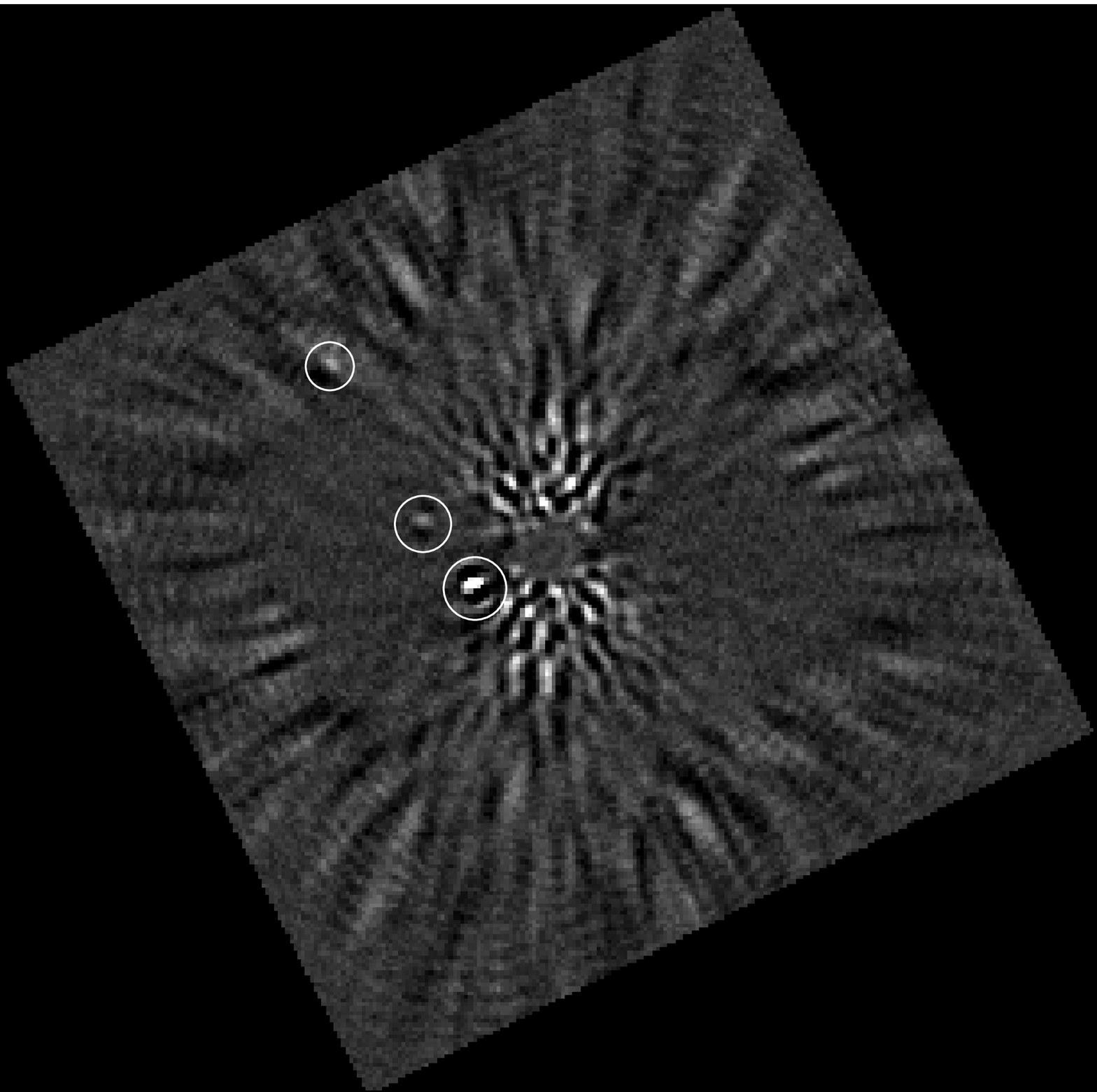}
	\end{center}
	\caption{\emph{Left}: Simulated 60 second single exposure using the Gemini Planet Imager instrument.  The bright spots are astrometric calibration spots generated by the instrument.  The bright lobes in the dark region are due to atmospheric turbulence and point along the major wind direction.  \emph{Center}: One hour of simulated data comprised of 60 second exposures, derotated and summed.  One planet is clearly visible with a second barely detectable.  \emph{Right}: PSF subtracted, summed version of the data set.  An additional planet becomes visible. \label{fig:gpi_data}}
\end{figure}

Multiple post-processing schemes have been proposed to improve the odds of extracting a planet signal.  In general, all of these attempt to model the point spread function (PSF) of the instrument, incorporating all static and quasi-static errors, and then subtract this (or decorrelate it) from the science image to reveal the residual planet signal.  This template PSF is constructed from data taken by the same instrument, but in which no planet signal is present in the same spatial location.  These data sets can either be historical libraries of other targets known not to have companions (or, at least, not to have companions that would appear in the same part of the image as the current target), or images of the same target star but with the planet signal appearing in different parts of the image plane.  The latter can be accomplished in a variety of ways---by producing angularly differentiated data by turning off the telescope de-rotator on ground based instruments (or spinning space-based observatories in between exposures) \citep{marois2006angular}; or by simultaneously imaging in multiple wavelengths, as with an integral field spectrograph \citep{racine1999speckle}; or by imaging in multiple polarizations \citep{stam2004using}.  All of these techniques produce data sets with spatially correlated noise, and decorrelated signal.  Once a data set of this sort has been created, it is possible to model the underlying noise distribution and to generate a PSF template consistent with the data but not containing the planet signal we wish to extract.  Figure \ref{fig:gpi_data} presents a sample data set of this type, showing simulated data for the Gemini Planet Imager \citep{macintosh2012gemini}.  The first image shows a single 60 second exposure with high shot and speckle noise.  The second image shows the summation of one hour of such images, which reveal the static and quasi-static elements of the noise.  The third image demonstrates the residuals after PSF subtraction, revealing the three planet signals in the original data set.

In \S\ref{sec:sig_extract}, we present a standardized notation and problem statement and briefly derive some of the most commonly used post-processing techniques, demonstrating the ubiquity of the image-to-image covariance calculation.  In \S\ref{sec:est}, we discuss how to most efficiently calculate the covariance and its inverse, using both sequential calculations and deriving a method to account for small changes in the reference image set.  We conclude with a sample implementation highlighting the utility of these techniques. 

\section{Signal Extraction}\label{sec:sig_extract}

\subsection{Problem Statement and Notation}\label{sec:prob_statement}
We assume that we have a set of reference images containing random realizations drawn from some distribution of noise.  We also have a target image containing both a noise component drawn from the same distribution, as well as the signal we wish to discover.  Frequently, the reference and target images are drawn from the same data set, with the target signal appearing in different spatial locations throughout the data.  Given this set of reference images and the target image, we wish to construct an estimate that minimizes the distance from the reference set and maximizes the distance from the target image in the spatial location of the planet signal (thereby minimizing the noise).  We write our set of $n$ vectorized reference images of dimension $p$ as  $\{\mf r_i\}_{i=1}^n$, where $\mf r_i \in \Re^{p}$, and the (vectorized) target image as $\mf t$.  The ordering of the vectorization is arbitrary, save that it be applied in the same way to each image (i.e., the final vectors can be stacked image columns, or transposed, stacked image rows, or any other consistent method of reforming a 2D image of $p$ elements into a column vector of dimension $p$).  We will use an overbar to represent the vector-mean subtracted value of each vector:
\begin{equation}
\mfbar t = \mf t - \mu(\mf t) \,; \qquad {\mfbar r}_i = \mf r_i - \mu(\mf r_i) \,,
\end{equation}
where $\mu(\mf x) \triangleq (\sum_{i=1}^p(x_i))/p$ for a $p$-element vector $\mf x$ with components $x_1,x_2,\ldots x_p$.
We form a $n\times p$ matrix $R$ whose rows are the transposed elements of the set $\{\mf r_i\}$:
\begin{equation}
R = \left[ \mf r_1, \mf r_2, \ldots, \mf r_n \right]^T \,,
\end{equation}
and the analgous row-mean subtracted matrix:
\begin{equation}\label{eq:rbar}
\bar R = \left[  {\mfbar r}_1, {\mfbar r}_2, \ldots,  {\mfbar r}_n \right]^T \,.
\end{equation}
Therefore the (image-to-image) sample covariance of the reference set is given by:
\begin{equation}
S \triangleq \frac{1}{p-1} \bar R \bar R^T \,,
\end{equation}
where $S$ has dimension $n \times n$.  The pixel to pixel covariance is thus:
\begin{equation}\label{eq:sprime}
S' \triangleq \frac{1}{n-1} \bar R^T \bar R \,,
\end{equation}
and has dimension $p \times p$.  

\subsection{Locally Optimal Combination of Images}
One method of solving the stated problem is to generate a least-squares fit to the target image via a linear combination of the reference set.  The first implementation of this method for exoplanet imaging was described in \citet{lafreniere2007new} as Locally Optimal Combination of Images (LOCI).  Written in the formalism of \S\ref{sec:prob_statement}, this approach requires finding the $n$-dimensional vector of optimal coefficients, $\mf c$, such that:
\begin{equation}\label{eq:minc}
\mf c = \arg \min_{\mf c} \left\Vert \mf t - R^T \mf c \right\Vert
\end{equation}
where  the norm is typically $\ell^2$. 

The LOCI procedure is analogous to solving the overdetermined linear system:
\begin{equation}\label{eq:csetup}
R^T \mf c = \mf t \,.
\end{equation}
As there are typically more pixels than independent images (i.e., $p > n$), a unique solution will not exist for \refeq{eq:csetup}.  However, when $RR^T$ is full rank, the left-pseudo-inverse of $R^T$ gives the minimum least-squares error solution: 
\begin{equation}\label{eq:pinv}
\mf c = (RR^T)^{-1}R\mf t \,,
\end{equation}
which satisfies \refeq{eq:minc} for the $\ell^2$ norm.  The case where $RR^T$ is not directly invertible is treated in the next section.

The estimated signal is the subtraction of the linear combination of references from the target image:
\begin{equation}\label{eq:loci_signal1}
\mfhat s = \mf t - R^T \mf c = (I - R^T(RR^T)^{-1}R)\mf t
\end{equation}
where $I$ is the identity matrix.  It is important to remember that \refeq{eq:pinv} is not an exact solution for \refeq{eq:csetup}, but rather a solution to \refeq{eq:minc}.  If an exact solution existed, this would imply that the signal is in the image of the reference set---that is, a linear combination of the noise---and we would get a zero signal estimate, making this method inappropriate for this task.

We can also perform the same least-squares fitting to the target image using the zero-mean reference set ($\bar R$) rather than the original reference set ($R$).    All of the steps in Equations (\ref{eq:minc})-(\ref{eq:loci_signal1}) remain the same, and our signal estimate is now:
\begin{equation}\label{eq:loci_signal2}
\mfhat s  = \left(I - \bar R^T\frac{S^{-1}}{p-1}\bar R\right)\mfbar t \,.
\end{equation}

The signal estimates in Equations (\ref{eq:loci_signal1}) and (\ref{eq:loci_signal2}) are not equal. In particular, \refeq{eq:loci_signal2} generates a zero-mean estimate (i.e., $\mu(\mfhat{s}) = 0$) whereas \refeq{eq:loci_signal1} generates a vector with mean proportional to the difference between the sample means of the target and references and the underlying distribution mean.  The operator in \refeq{eq:loci_signal2} is also approximately mean and norm preserving, meaning that if it is applied to $\mf t$ rather than $\mfbar t$, the resulting signal estimate would have approximately the same mean as the target image, with the variance of the signal estimate for the mean-subtracted target image.  From the definition of the covariance, we see that $S$ can be written as:
\begin{equation}
S = \frac{1}{p-1}\left(RR^T - p \boldsymbol{\mu}_R\boldsymbol{\mu}_R^T\right)
\end{equation}
where $\boldsymbol{\mu}_R$ is the vector of row means of $R$:
\begin{equation}
\boldsymbol{\mu}_R \triangleq \frac{R}{p}  \mf 1_{p\times 1} \,,
\end{equation}
and $ \mf 1_{p\times 1}$ is the $p$ element column vector of ones.  By the Sherman-Morrison formula \citep{sherman1950adjustment}, this means that 
\begin{equation}\label{eq:sherman1}
S^{-1} = (p-1)\left[(RR^T)^{-1} + p\frac{(RR^T)^{-1} \boldsymbol{\mu}_R\boldsymbol{\mu}_R^T (RR^T)^{-1}}{1 - p\boldsymbol{\mu}_R^T(RR^T)^{-1}\boldsymbol{\mu}_R}\right] \,,
\end{equation}
and
\begin{equation}\label{eq:sherman1}
(RR^T)^{-1} = \frac{1}{p-1}\left[S^{-1} - \frac{p}{p-1}\left(\frac{S^{-1} \boldsymbol{\mu}_R\boldsymbol{\mu}_R^T S^{-1}}{1 + \frac{p}{p-1}\boldsymbol{\mu}_R^TS^{-1}\boldsymbol{\mu}_R}\right)\right] \,.
\end{equation}
The second term in \refeq{eq:sherman1} scales as $p/(p-1)^2$ and so goes to zero in the limit as $p \rightarrow \infty$.  For finite matrix sizes, the difference between $S^{-1}$ and $(p-1)(RR^T)^{-1}$ is a small value proportional to the difference between the calculated sample means and the true means of the underlying distribution from which the references and target are sampled.  The zero-mean properties of \refeq{eq:loci_signal2} make it easier to extract unbiased estimates of the scene, so we will use this form going forward.

\subsection{Covariance Pseudoinverses}
Of course, $S$ is only guaranteed to be positive semi-definite and is therefore not necessarily invertible ($S$ is only positive definite when all rows of $R$ are linearly independent).   In these cases we can use pseudoinverses of the covariance to calculate the signal estimates.  One option is to use singular value decomposition (SVD) based inversion as in \citet{marois2010exoplanet}, which describes this technique as part of the Speckle-Optimize Subtraction for Imaging Exoplanets (SOSIE) pipeline.  Real matrix $S$ can be decomposed as
\begin{equation}
S = U \Sigma V^T \,,
\end{equation}
where $\Sigma$ is a positive semi-definite diagonal matrix of singular values and $U$ and $V$ are unitary.  Because $S$ is square, all of these matrices will be square and of the same dimensions as $S$. The pseudoinverse of $S$ can then be expressed as:
\begin{equation}
S^+ = V \Sigma^+U^T \,,
\end{equation}
where $\Sigma^+$ is the pseudoinverse of $\Sigma$---non-zero entries of $\Sigma$ are replaced by their reciprocals while zero (or small) values are left as zeros.  Assuming that the diagonal of $\Sigma$ is ordered by decreasing magnitude (these values can always be reordered as the columns of $U$ and $V$ form orthogonal bases for the left and right singular vectors of $S$), this can be expressed as:
\begin{equation}\label{eq:sigmapseudo}
\Sigma^+ = \begin{bmatrix}
I_{P\times P} & \mf 0_{P \times (n-P)} \\
\mf 0_{(n-P)\times P} & \mf 0_{(n - P) \times (n-P)} 
\end{bmatrix} \Sigma^{-1} 
\end{equation}
where $P$ is the number of singular values (diagonals of $\Sigma$) that are greater than a desired tolerance,  $\epsilon$, and where $\mf 0_{m\times n}$ represents an $m \times n$ matrix of zeroes and $I_{n\times n}$ represents an $n \times n$ identity matrix.
\refeq{eq:loci_signal2} thus becomes:
\begin{equation}
\mfhat s  = \left(I - \frac{1}{N-1} \bar R^T V \Sigma^+ U^T \bar R\right)\mfbar t \,.
\end{equation}

Alternatively, we can use eigendecomposition, which, in this case, is mathematically equivalent to SVD.  $S$ is Hermitian (symmetric in the strictly real case) and therefore is diagonizable as:
\begin{equation}\label{eq:Phidef}
S\Phi = \Phi \Lambda \,,
\end{equation}
where $\Phi$ is the unitary $n\times n$ matrix  whose columns are the eigenvectors of $S$ and form an orthonormal basis, and $\Lambda$ is a diagonal $n \times n$ matrix whose entries are the corresponding eigenvalues.  We assume that $\Phi$ and $\Lambda$ are ordered such that the eigenvalues decrease from largest to smallest along the diagonal of $\Lambda$.  Thus,
\begin{equation}
S^{-1} = \Phi \Lambda^{-1} \Phi^T \,,
\end{equation}
where the pseudoinverse of $\Lambda$ may be used in cases of small or zero eigenvalues to find the pseudoinverse of $S$.  This is calculated in the same manner as the pseudoinverse of $\Sigma$ in \refeq{eq:sigmapseudo}.

\subsection{Karhunen-Lo\`{e}ve Image Processing}
While the previous section describes viable methods for regularizing the covariance and achieving an inverse calculation, there is another possible approach:  Rather than using a pseudoinverse, we can project the target signal onto a subset of an optimally energy compacting basis using the Karhunen-Lo\`{e}ve (KL) theorem, as in \citet{soummer2012detection}.  To do so, we define:
\begin{equation}
Z \triangleq \Phi^T \bar R \,,
\end{equation}
where $Z$ is the $n \times p$ matrix of KL transform vectors and $\Phi$ is the matrix defined by \refeq{eq:Phidef}.  

From our earlier definition of the reference set, we know that the matrix $ZZ^T$ will be positive semi-definite in the general case (and positive-definite when all elements of the reference set are linearly independent), and thus has a unique principal square root \citep{horn2012matrix}.  Using the shorthand $B= \sqrt{A} \iff BB=A$ for the matrix square root, we can write:
\begin{equation}
\begin{split}
\sqrt{Z Z^T} &= \sqrt{\Phi^T \bar R (\Phi^T \bar R)^T} \\
&= (\sqrt{p-1})\sqrt{\Phi^T S \Phi} =  (\sqrt{p-1})\sqrt{\Lambda}  \,.
\end{split}
\end{equation}

Defining the diagonal matrix G (where again the pseudo-inverse of $\Lambda$ can be used in cases of zero eigenvalues of $S$) as:
\begin{equation}
G \triangleq \frac{1}{\sqrt{p-1}} \sqrt{\Lambda^{-1}} \,,
\end{equation}
we can write the zero row mean, normalized version of $Z$ as:
\begin{equation}
\bar Z \triangleq G Z \,.
\end{equation}
The matrix $\bar Z$ is our linear transform operator and is a decorrellating, optimally compacting basis for $\bar R$ \citep{rao2000transform}.  In order to drop any zero eigenvalues in the case where the reference set elements are not linearly independent, and to avoid overfitting the target, $\bar Z$ is truncated to $k$ rows:
\begin{equation}
\bar Z_k = I_k \bar Z \,,
\end{equation}
where $I_k$ is the $n\times n$ identity matrix truncated to the first $k$ rows (final dimension $k \times n$):
\begin{equation}
I_k = \begin{bmatrix}
 I_{k\times k}  & \mf 0_{k\times (n-k)} \end{bmatrix} \,.
\end{equation}

To reconstruct the target image $\mf t$ we project it onto $\bar Z_k$:
\begin{equation}
\hat{\mf t} = \bar Z_k^T\bar Z_k \mf t \,,
\end{equation}
and, as before, recover the signal by subtracting this from the original image:
\begin{equation}
\mfhat s  = \left(I - \bar Z_k^T\bar Z_k \right)\mfbar t \,.
\end{equation}

This is equivalent to:
\begin{equation}
\mfhat s  = \left(I - \frac{1}{p-1} \bar R^T\Phi \Lambda^{-1} F  \Phi^T \bar R \right)\mfbar t
\end{equation}
where $F$ is the matrix formed by padding the transpose of $I_k$ with $n - k$ zero columns:
\begin{equation}
F = \begin{bmatrix} I_{k\times k}  & \mf 0_{k\times (n-k)} \\ \mf 0_{ (n-k) \times k}  & \mf 0_{(n-k)\times (n-k)}   \end{bmatrix}\,.
\end{equation}
This procedure has been titled Karhunen-Lo\`{e}ve Image Processing (KLIP) \citep{soummer2012detection}.

\subsection{Hotelling Observer}
Finally, we have the case of the Hotelling Observer \citep{barrett2006objective,caucci2007application}, the optimal linear discriminant whose test statistic is calculated via the inverse of the full data covariance.  For a single image set, in our notation, this statistic would be:
\begin{equation}
\left((S')^{-1}\mfhat s \right)^T \mf t
\end{equation}
where $S'$ is the pixel to pixel covariance defined in \refeq{eq:sprime} (c.f., \citet{caucci2007application} Equations 16-17). 

As the dimensionality of the full data covariance is much larger than that of the image-to-image covariance for typical data sets, a direct inversion may be more difficult, and significantly more data is required for the matrix to be well conditioned \citep{lawson2012advanced}. This has led to multiple proposed decompositions for the data covariance, with some factors estimated via simulated data, and certain simplifying assumptions including exact knowledge of the background and full statistical knowledge of the PSF as in \citet{caucci2007application}.  Implementations of these calculations have been successfully demonstrated on specialized high-performance computing environments \citep{caucci2009spatio}.

\section{Sequential and Neighboring Calculations}\label{sec:est}

All of the techniques described in the previous section make heavy use of the covariance of the reference data set.  For the  LOCI and KLIP-like algorithms, it is often necessary to calculate hundreds of covariance matrices of reference sets containing many of the same images.  This is especially true when using data sets derived from angular or spectral diversity, where the reference set for each individual image is the remainder of the data set, minus a small number of images where the planet signal would occur in the same general location as in the target image.  For Hotelling observers and library-based templates we wish to calculate large covariance matrices possibly including all of the images taken by an instrument to date, which can be a very time and memory intensive operation.  Finally, when applying region-based implementations of KLIP and LOCI we may wish to make small modifications to the regions used to optimize the processes (see \citet{marois2010exoplanet} and \citet{pueyo2012application} for detailed discussions on LOCI optimization zones).  In each case, we can greatly improve the efficiency of our calculations by replacing batch and redundant processes with sequential ones (see \citet{stengel1994optimal} for a general discussion on sequential processing techniques).

\subsection{Mean and Covariance Update}
As a first step in developing the tools specific to our application, we will outline the sequential calculation methods for finding the sample mean and covariance of a data set. Given a set of vectors $\{\mf x_i\}_1^n$ we define the sample mean as
\begin{equation}\label{eq:mean}
\bmu = \frac{1}{n}\sum_{i=1}^n \mf x_i \,,
\end{equation}
and sample covariance as
\begin{equation}\label{eq:cov}
\mf S = \frac{1}{n-1}\sum_{i=1}^n (\mf x_i - \bmu)(\mf x_i -\bmu)^T \,.
\end{equation}
Expanding the summation in \refeq{eq:cov}, we have
\begin{equation}\label{eq:covexpand}
\begin{split}
S &= \frac{1}{n-1}\sum_{i=1}^n \left[ \mf x_i \mf x_i^T - \bmu \mf x_i^T - \mf x_i \bmu^T + \bmu \bmu^T \right] \\
&= \frac{1}{n-1}\left[ \sum_{i=1}^n \mf x_i \mf x_i^T -  \bmu \sum_{i=1}^n \mf x_i^T - \sum_{i=1}^n \mf x_i \bmu^T + n\bmu \bmu^T \right]\\
&= \frac{1}{n-1}\left[ \sum_{i=1}^n \mf x_i \mf x_i^T -  \bmu n\bmu^T - n\bmu\bmu^T + n\bmu \bmu^T \right]\\
&= \frac{1}{n-1}\left[ \sum_{i=1}^n \mf x_i \mf x_i^T -  n\bmu \bmu^T \right] \,,
\end{split}
\end{equation}
where the penultimate step is due to the definition of the mean from \refeq{eq:mean}.

Now, let the mean and covariance at time $k$ be denoted by $\bmu_k$ and $\mf S_k$.  Then:
\begin{align}
(k-1)\bmu_{k-1} &=  \sum_{i=1}^{k-1} \mf x_i \\
k\bmu_{k} &=  \sum_{i=1}^{k} \mf x_i \,,
\end{align}
so that
\begin{align}
k\bmu_{k} - (k-1)\bmu_{k-1} &=  \sum_{i=1}^{k} \mf x_i - \sum_{i=1}^{k-1} \mf x_i  = \mf x_k \Rightarrow\\
\bmu_k &= \frac{(k-1)\bmu_{k-1} + \mf x_k}{k} \label{eq:muupdate}\,.
\end{align}
Similarly, 
\begin{align}
(k-2)\mf S_{k-1} &=  \sum_{i=1}^{k-1} \mf x_i \mf x_i^T -  (k-1)\bmu_{k-1} \bmu^T_{k-1}\label{eq:skm1}\\
(k-1)\mf S_{k} &=  \sum_{i=1}^{k} \mf x_i \mf x_i^T -  k \bmu_{k} \bmu^T_{k}\,,
\end{align}
so
\begin{equation}
(k-1)\mf S_{k} - (k-2)\mf S_{k-1} = \sum_{i=1}^{k} \mf x_i \mf x_i^T - \sum_{i=1}^{k-1} \mf x_i \mf x_i^T -  k \bmu_{k} \bmu^T_{k} + (k-1)\bmu_{k-1} \bmu^T_{k-1} \,.
\end{equation}
Substituting $\bmu_{k-1}$ with the expression derived from \refeq{eq:muupdate}, this becomes
\begin{align}
(k-1)\mf S_{k} &=  (k-2)\mf S_{k-1} +  \mf x_k \mf x_k^T -  k \bmu_{k} \bmu^T_{k} + (k-1)\left[\left(\frac{k\bmu_k - \mf x_k}{k-1}\right)\left(\frac{k\bmu_k^T - \mf x_k^T}{k-1}\right)\right] \Rightarrow\\
\mf S_{k} &=  \frac{k-2}{k-1}\mf S_{k-1} +  \frac{k}{(k-1)^2}\left[ (\mf x_k -  \bmu_{k})(\mf x_k - \bmu_k)^T\right] \label{eq:covupdate}\,.
\end{align}

Alternatively, from \refeq{eq:cov}, we can write
\begin{equation}
(k-1)\mf S_k = \sum_{i=1}^{k-1} (\mf x_i - \bmu_k)(\mf x_i -\bmu_k)^T +  (\mf x_k - \bmu_k)(\mf x_k -\bmu_k)^T
\end{equation}
so the final term in \refeq{eq:covupdate} becomes
\begin{equation}\label{eq:buh}
(\mf x_k -  \bmu_{k})(\mf x_k - \bmu_k)^T = (k-1)\mf S_k - \sum_{i=1}^{k-1} (\mf x_i - \bmu_k)(\mf x_i -\bmu_k)^T \,.
\end{equation}
Substituting \refeq{eq:muupdate} for $\bmu_k$, the summation in the above equation becomes
\begin{equation}
\begin{split}
\sum_{i=1}^{k-1} &(\mf x_i - \bmu_k)(\mf x_i -\bmu_k)^T \\
 =&\sum_{i=1}^{k-1} \mf x_i \mf x_i^T - \frac{k-1}{k} \bmu_{k-1}\sum_{i=1}^{k-1} \mf x_i^T - \frac{1}{k} \mf x_k \sum_{i=1}^{k-1} \mf x_i^T - \frac{k-1}{k}\sum_{i=1}^{k-1} \mf x_i  \bmu_{k-1}^T - \frac{1}{k} \sum_{i=1}^{k-1} \mf x_i  \mf x_k^T\\
 &{} + \sum_{i=1}^{k-1}\left[ \frac{(k-1)^2}{k^2}\bmu_{k-1}\bmu_{k-1}^T + \frac{k-1}{k^2}\bmu_{k-1}\mf x_k^T + \frac{k-1}{k^2}\mf x_k\bmu_{k-1}^T + \frac{1}{k^2}\mf x_k \mf x_k^T \right]\\
 =&\sum_{i=1}^{k-1} \mf x_i \mf x_i^T + \left(\frac{k-1}{k^2} - (k-1)\right) \bmu_{k-1}\bmu_{k-1}^T - \frac{k-1}{k^2}\left(\bmu_{k-1}\mf x_k^T + \mf x_k\bmu_{k-1}^T - \mf x_k \mf x_k^T\right)\\
 =& (k-2) \mf S_{k-1} + \frac{k-1}{k^2}\left(\bmu_{k-1}\bmu_{k-1}^T  - \bmu_{k-1}\mf x_k^T - \mf x_k\bmu_{k-1}^T + \mf x_k \mf x_k^T\right) \,,
\end{split}
\end{equation}
where we used \refeq{eq:skm1} in the final step.  Returning to \refeq{eq:buh}, we can now write
\begin{equation}
(\mf x_k -  \bmu_{k})(\mf x_k - \bmu_k)^T = (k-1)\mf S_k - (k-2) \mf S_{k-1} - \frac{k-1}{k^2}\left(\bmu_{k-1}\bmu_{k-1}^T  - \bmu_{k-1}\mf x_k^T - \mf x_k\bmu_{k-1}^T + \mf x_k \mf x_k^T\right) \,.
\end{equation}
Substituting this back into \refeq{eq:covupdate} yields
\begin{equation}
\mf S_{k} =  \frac{k-2}{k-1}\mf S_{k-1} +  \frac{1}{k}\left[(\mf x_k - \bmu_{k-1})(\mf x_k - \bmu_{k-1})^T\right] \label{eq:covupdate2}\,.
\end{equation}

Both versions of the covariance update (\refeq{eq:covupdate} and \refeq{eq:covupdate2}) may be written as
\begin{equation}\label{eq:covupdates}
\mf S_{k} = \alpha\mf S_{k-1} + \beta \mf v_k \mf v_k^T \,,
\end{equation}
with 
\begin{equation*}
\begin{tabular}{ccc}
$\alpha$ & $\beta$ &$\mf v_k$\\
\hline
 $\frac{k-2}{k-1} $ & $\frac{k}{(k-1)^2}$ & $\mf x_k - \bmu_k$\\ 
 $\frac{k-2}{k-1} $ & $1/k$ & $\mf x_k - \bmu_{k-1}$
\end{tabular}
\end{equation*}

\subsection{Square Root and Inverse Updates}
In several applications in \S\ref{sec:sig_extract} we are interested in the inverse of the covariance more so than the covariance itself.  Fortunately, there are known, simple matrix decompositions that allow us to update the inverse covariance directly with new sample vectors, rather than updating the covariance and recalculating the inverse at each step. As in \citet{igel2006computational}, it can be shown that the update of the Cholesky decomposition of the covariance,
\begin{equation}
\mf S = \mf L \mf L^T
\end{equation}
where $\mf L$ is a lower triangular matrix with positive diagonal values, can be written as
\begin{equation}
\mf L_{k} = \sqrt{\alpha}\mf L_{k-1} + \frac{\sqrt{\alpha}}{\Vert \mf z_k \Vert^2} \left(\sqrt{1 + \frac{\beta}{\alpha}\Vert \mf z_k \Vert^2} - 1 \right) \mf v_k\mf z_k^T \,,
\end{equation}
where $\mf z_k$ is the vector defined implicitly by
\begin{equation}\label{eq:zdef}
\mf v_k = \mf L_{k-1}\mf z_k
\end{equation}
so that 
\begin{equation}
\Vert \mf z_k \Vert^2 = \mf v_k^T \mf L_{k-1}^{-T}\mf L_{k-1}^{-1} \mf v_k \,,
\end{equation}
where $()^{-T}$ represents the inverse-transpose.
Thus, rather than updating the full covariance, we can update its square root (in the Cholesky sense), which is a potentially more useful quantity.  However, this approach still requires that we calculate the inverse of $\mf L$ at each time step, whereas we wish to update the inverse covariance, or some decomposition of it, instead.  

By the Sherman-Morrison formula \citep{sherman1950adjustment}
\begin{equation}\label{eq:sinvupdate}
\mf S_k^{-1} = \frac{\mf S_{k-1}^{-1} }{\alpha} -  \frac{\mf S_{k-1}^{-1}}{\alpha}\left(I + \frac{\beta}{\alpha}\mf v_k \mf v_k^T \mf S_{k-1}^{-1}\right)^{-1}\frac{\beta}{\alpha}\mf v_k \mf v_k^T \mf S_{k-1}^{-1} \,,
\end{equation}
where $I$ is the identity matrix. Note that in this formulation, we never need to invert $\mf v_k \mf v_k^T$ itself, avoiding any ill-conditioning problems.  However, in numerical applications, it is important to always evaluate $\mf v_k \mf v_k^T$ first, before any other matrix multiplications, so as to guarantee that numerical errors do not corrupt the Hermitian property of the resulting matrix. Substituting
\begin{equation}
\mf S^{-1} =  \mf L^{-T} \mf L^{-1} \,,
\end{equation}
\refeq{eq:sinvupdate} becomes
\begin{equation}\label{eq:linvupdate1}
\begin{split}
 \mf L_k^{-T} \mf L_k^{-1} &= \frac{ \mf L_{k-1}^{-T} \mf L_{k-1}^{-1} }{\alpha} -  \frac{ \mf L_{k-1}^{-T} \mf L_{k-1}^{-1}}{\alpha}\left(I + \frac{\beta}{\alpha}\mf v_k \mf v_k^T  \mf L_{k-1}^{-T} \mf L_{k-1}^{-1}\right)^{-1}\frac{\beta}{\alpha}\mf v_k \mf v_k^T \mf L_{k-1}^{-T} \mf L_{k-1}^{-1}\\
 &= \frac{ \mf L_{k-1}^{-T} }{\sqrt{\alpha}}\left[ I - \frac{\beta}{\alpha}\mf L_{k-1}^{-1} \left(I + \frac{\beta}{\alpha}\mf v_k \mf v_k^T  \mf L_{k-1}^{-T} \mf L_{k-1}^{-1}\right)^{-1}\mf v_k \mf v_k^T \mf L_{k-1}^{-T} \right] \frac{ \mf L_{k-1}^{-1} }{\sqrt{\alpha}} \,.
\end{split}
\end{equation}
Again by the Sherman-Morrison formula, the bracketed term above can be rewritten as
\begin{equation}
I - \frac{\beta}{\alpha}\mf L_{k-1}^{-1} \left(I + \frac{\beta}{\alpha}\mf v_k \mf v_k^T  \mf L_{k-1}^{-T} \mf L_{k-1}^{-1}\right)^{-1}\mf v_k \mf v_k^T \mf L_{k-1}^{-T} = \left(I + \frac{\beta}{\alpha} \mf L_{k-1}^{-1} \mf v_k \mf v_k^T  \mf L_{k-1}^{-T} \right)^{-1} \,.
\end{equation}
Defining matrix $\mf T$ and its Cholesky decomposition, $\mf M$, as:
\begin{equation}
\mf T_k \triangleq I + \frac{\beta}{\alpha} \mf L_{k-1}^{-1} \mf v_k \mf v_k^T  \mf L_{k-1}^{-T} \quad\textrm{and}\quad \mf T_k = \mf M_k \mf M_k^T\,,
\end{equation}
\refeq{eq:linvupdate1} becomes
\begin{equation}
\mf L_k^{-T} \mf L_k^{-1} =  \frac{ \mf L_{k-1}^{-T} }{\sqrt{\alpha}} \left(\mf M_k^{-T} \mf M_k^{-1}\right)\frac{ \mf L_{k-1}^{-1} }{\sqrt{\alpha}} \,,
\end{equation}
so the update of the inverse Cholesky factor is simply
\begin{equation}
\mf L_k^{-1} = \frac{1}{\sqrt{\alpha}} \mf M_k^{-1} \mf L_{k-1}^{-1} \,.
\end{equation}
Note that many of the inverses in the expressions above need not be directly calculated, but can be replaced with the equivalent, specialized LAPACK routines for solving systems of linear equations \citep{anderson1999lapack}. All terms of the form $A^{-1}B$ are the solution of the linear system $AX=B$, and have multiple dedicated solvers, the choice of which depends on the specific form of $A$ and $B$. 

\subsection{Initialization}
In cases where $n$ is smaller than the dimension of $\mf x$ (as in the initial stages of collecting a large data set) , the covariance is not full rank and so the inverse covariance is undefined.  Furthermore, the covariance in these cases is not guaranteed to be positive definite so that the Cholesky factor will also be undefined.  Even if the final number of samples is greater than the size of each sample vector, this condition will persist in the initialization and early updates of the covariance.  

To address this, we can use an indefinite decomposition (closely related to the Cholesky decomposition - see \citet{watkins2004fundamentals} for details), such as 
\begin{equation}
\mf S = \blam \mf D \blam^T
\end{equation}
where $\blam$ is again a lower triangular matrix, while $\mf D$ is a diagonal matrix.  Then, \refeq{eq:covupdates} becomes
\begin{equation}
 \blam_k \mf D_k \blam^T_k = \alpha\mf  \blam_{k-1} \mf D_{k-1} \blam^T_{k-1} + \beta \blam_{k-1} \mf z_k \mf z_k^T \blam_{k-1}^{T} \,,
\end{equation}
where $\mf z_k$ is again defined via
\begin{equation}\label{eq:zdef2}
\mf v_k = \blam_{k-1}\mf z_k\,.
\end{equation}
Therefore, the factors of this decomposition may be updated as 
\begin{align}
\blam_k &= \sqrt{\alpha} \blam_{k-1}\mf M_k\\
\mf D_k &= \mf G_k \,,
\end{align}
where
\begin{equation}
\mf M_k \mf G_k \mf M_k^T \triangleq \mf D_{k-1} + \beta \mf z_k \mf z_k^T \,.
\end{equation}
With this definition, the inverse of $\blam_k$ can always be found, even when $\mf S_k$ is singular.  As the update progresses to a point where $\mf S_k$ is positive definite, the diagonal elements of $\mf D_k$ will all become positive.  At this point, we can convert to the Cholesky factor as
\begin{equation}
\mf L = \blam \sqrt{\mf D} \,.
\end{equation}

\subsection{Cross-Covariance}
In several of these applications, we will want to also produce covariances conditioned on some other set of available information not included in the images themselves.  For example, we may want to account for atmospheric conditions, or instrument operating conditions, etc. (see \citet{caucci2009spatio} for details).  This conditioning is provided by the calculation of the cross-covariance, which can also be updated sequentially.  Given a second set of vectors $\{\mf y_i\}_1^n$, the cross-covariance with $\{\mf x_i\}_1^n$ is 
\begin{equation}
\leftexp{\mf{xy}}{\mf S} = \frac{1}{n-1}\sum_{i=1}^n (\mf x_i - \leftexp{\mf x}{\bmu})(\mf y_i - \leftexp{\mf y}{\bmu})^T
\end{equation}
where $ \leftexp{\mf x}{\bmu}$ and $ \leftexp{\mf y}{\bmu}$ are now the means of the two sets, respectively.  As before, it can be shown that
\begin{equation}
\leftexp{\mf{xy}}{\mf S} = \frac{1}{n-1}\left[ \sum_{i=1}^n \mf x_i \mf y_i^T -  n \leftexp{\mf x}{\bmu}  \leftexp{\mf y}{\bmu}^T \right] 
\end{equation}
and 
\begin{equation}
\leftexp{\mf{xy}}{\mf S}_{k} =  \frac{k-2}{k-1}\leftexp{\mf{xy}}{\mf S}_{k-1} +  \frac{k}{(k-1)^2}\left[ (\mf x_k -  \leftexp{\mf x}{\bmu}_{k})(\mf y_k - \leftexp{\mf y}{\bmu}_k)^T\right] 
\end{equation}
We now have 
\begin{equation}
\leftexp{\mf{xy}}{\mf S}_{k} = \alpha\leftexp{\mf{xy}}{\mf S}_{k-1}  + \beta \mf v_k \mf w_k^T \,,
\end{equation}
with all values defined as before, and $\mf w_k = \mf y_k -  \leftexp{\mf y}{\bmu}_{k}$ or $\mf y_k -  \leftexp{\mf y}{\bmu}_{k-1}$.  Again defining the Cholesky factor of $\leftexp{\mf{xy}}{\mf S}$ as
\begin{equation}
\leftexp{\mf{xy}}{\mf S} = \leftexp{\mf{xy}}{\mf L} \leftexp{\mf{xy}}{\mf L}^T \,,
\end{equation}
the Cholesky factor updates are now
\begin{align}
 \leftexp{\mf{xy}}{\mf L}_{k} &= \sqrt{\alpha} \leftexp{\mf{xy}}{\mf L}_{k-1}\mf M_k \\
 \leftexp{\mf{xy}}{\mf L}_{k}^{-1} &= \frac{1}{\sqrt{\alpha}} \mf M_k^{-1} \leftexp{\mf{xy}}{\mf L}_{k-1}^{-1}
\end{align}
where
\begin{equation}
\left(I + \frac{\beta}{\alpha}  \leftexp{\mf{xy}}{\mf L}^{-1}_{k-1} \mf v_k \mf w_k^T \leftexp{\mf{xy}}{\mf L}^{-T}_{k-1} \right) = \mf M_k \mf M_k^T \,.
\end{equation}

\subsection{Neighboring Covariances}
Finally, there is the case of overlapping subsets of vectors for which we wish to calculate covariances.  In the cases of KLIP and LOCI, we will frequently wish to update the covariance with one or more new reference images, but also to remove one or more images from the reference set.  A concrete illustration of this is the case of applying the algorithm to an angular diversity data set, where the noise in each image remains nearly static, while the planet signal moves about the center of rotation.  The reference set for each image in the data set is the subset of images where the planet signal is a minimum angular distance away from its location in the target image.   Thus, reference sets will be highly overlapping, with a relatively small number of images added or dropped in each subsequent reference set.  This is equivalent to calculating the covariance for a matrix $\bar R$ with one or more rows removed.  

As demonstrated in \citet{lafreniere2007new} and elsewhere, removing rows from the $\bar R$ matrix is equivalent to simply removing rows and columns from the $S$ matrix calculated from the original $\bar R$.  For example, we can represent the removal of the $i$th row of $\bar R$ to form the truncated matrix $\bar R_{-i}$ as $\bar R_{-i} = H \bar R$, where $H$ is a block diagonal identity matrix with all zeros in the $i$th column:
\begin{equation}
H \triangleq \begin{bmatrix} 
\begin{matrix} I_{(i-1)\times (i-1)} \\ \mf 0_{ (n-i) \times (i-1)} \end{matrix} & 
 \begin{matrix} \mf 0_{(i-1)\times 1} \\ \mf 0_{ (n-i) \times 1} \end{matrix} & 
 \begin{matrix}  \mf 0_{((i-1) \times (n-i)} \\ I_{ (n-i) \times (n-i)} \end{matrix} \end{bmatrix} \,,
\end{equation}
with $\mf 0_{m\times n}$ representing an $m \times n$ matrix of zeroes and $I_{n\times n}$ the $n \times n$ identity matrix.
The associated covariance would then equal:
\begin{equation}
S_{-i} =  \frac{1}{p-1} H \bar R \bar R^T H^T \,,
\end{equation}
which is simply the removal of the $i$th row and column from $S$.

A less trivial case occurs when we wish to calculate the covariance of $\bar R$ having added or removed one or more \emph{columns} - equivalent to adding and subtracting pixel locations in the image.  This can be useful when you are working on optimizing zones in LOCI (see \citet{marois2010exoplanet} for discussion) or applying KLIP in varying or overlapping annular regions.  It is also applicable to similar optimizations that can be attempted for the Hotelling observer, and can be used together with the sequential calculation of the Hotelling covariance described above.

In order to describe this situation, we now define two non-intersecting sets of vectors $\{\mf x_i\}_1^n$ and $\{\mf y_i\}_1^m$ with sample means $\bmu_{\mf x}$, $\bmu_{\mf y}$ and sample covariances $\mf S_{\mf x}, \mf S_{\mf y}$, respectively.   Returning to the definition in \refeq{eq:mean} we see that the total sample mean of the union of the two sets (represented by $\mf x \cup \mf y$) is:
\begin{equation}\label{eq:musum}
\bmu_{\mf x \cup \mf y} = \frac{1}{n+m}\left(n\bmu_{\mf x} + m\bmu_{\mf y}\right) \,.
\end{equation}

Similarly, from \refeq{eq:covexpand} we can write:
\begin{align}
\mf S_{\mf x} &= \frac{1}{n-1}\left[ \sum_{i=1}^n \mf x_i \mf x_i^T -  n\bmu_{\mf x}\bmu_{\mf x}^T \right] \\
\mf S_{\mf y} &= \frac{1}{m-1}\left[ \sum_{i=1}^m \mf y_i \mf y_i^T -  m\bmu_{\mf y}\bmu_{\mf y}^T \right] \\
\mf S_{\mf x \cup \mf y} &= \frac{1}{n+m-1}\left[ \sum_{i=1}^n \mf x_i \mf x_i^T + \sum_{i=1}^m \mf y_i \mf y_i^T -  (n+m)\bmu_{\mf x +\mf y}\bmu_{\mf x \cup \mf y}^T \right] \label{eq:covsum}\,.
\end{align}
Substituting the first two equations into the third yields:
\begin{equation}\label{eq:covplus}
\begin{split}
\mf S_{\mf x \cup \mf y} &= \frac{1}{n+m-1}\left[ (n-1) \mf S_{\mf x} + (m-1)\mf S_{\mf y} +n\bmu_{\mf x}\bmu_{\mf x}^T + m\bmu_{\mf y}\bmu_{\mf y}^T - \frac{1}{n+m}\left(n\bmu_{\mf x} + m\bmu_{\mf y}\right)\left(n\bmu_{\mf x} + m\bmu_{\mf y}\right)^T  \right] \\
&= \frac{1}{n+m-1}\left[ (n-1) \mf S_{\mf x} + (m-1)\mf S_{\mf y} + \frac{nm}{n+m}\left(\bmu_{\mf x} - \bmu_{\mf y}\right)\left(\bmu_{\mf x} - \bmu_{\mf y}\right)^T \right] \,.
\end{split}
\end{equation}
Rewriting \refeq{eq:covsum} in a slightly different way, and substituting in \refeq{eq:musum} we get:
\begin{equation}\label{eq:covminus}
\begin{split}
\mf S_{\mf y} &=  \frac{1}{m-1}\left[ (n+m-1) \mf S_{\mf x +\mf y} - (n-1)\mf S_{\mf x} -n\bmu_{\mf x}\bmu_{\mf x}^T - m\bmu_{\mf y}\bmu_{\mf y}^T +(n+m)\bmu_{\mf x +\mf y}\bmu_{\mf x+\mf y}^T  \right] \\
&= \frac{1}{m-1}\left[ (n+m-1) \mf S_{\mf x +\mf y} - (n-1)\mf S_{\mf x} - \frac{n(n+m)}{m}\left(\bmu_{\mf x} - \bmu_{\mf x \cup \mf y} \right)\left(\bmu_{\mf x} - \bmu_{\mf x \cup \mf y} \right)^T\right] \, .
\end{split}
\end{equation}
Equations (\ref{eq:covplus}) and (\ref{eq:covminus}) give us the ability to quickly update a covariance by adding and removing elements from the reference set without recalculating the entire covariance matrix, leading a significantly smaller number of  operations, especially when the total data set is significantly larger than the number of images added and dropped for each reference set.

In the case where only one vector is being added and removed at each iteration, we can write a single set of update equations:
\begin{align}
\bmu_{k+1} &= \bmu_k + \frac{1}{N-1}\left[\mf x_{k-1} - \mf x_{k}\right] \label{eq:meanupdate1} \\
\mf S_{k+1} &=  \mf S_{k} + \frac{N-1}{(N-2)^2}\left[ \left(\mf x_{k-1} - \bmu_{k+1}\right)\left(\mf x_{k-1} - \bmu_{k-1}\right)^T -   \left(\mf x_k - \bmu_{k}\right)\left(\mf x_k - \bmu_{k}\right)^T \right] \label{eq:covupdate1}
\end{align}
where a set of $N$ sample vectors is taken $N-1$ at a time, with vector $\mf x_k$ dropped and vector $\mf x_{k-1}$ added at each step.  Thus, $S_{k}$ and $\bmu_k$ are always the covariance and mean of the subset $\{\mf x_i\}_{i\ne k}$ of the sample vector set $\{\mf x_i\}_{i=1}^N$. 

\begin{figure}
	\begin{center}
	\includegraphics[width=1\textwidth]{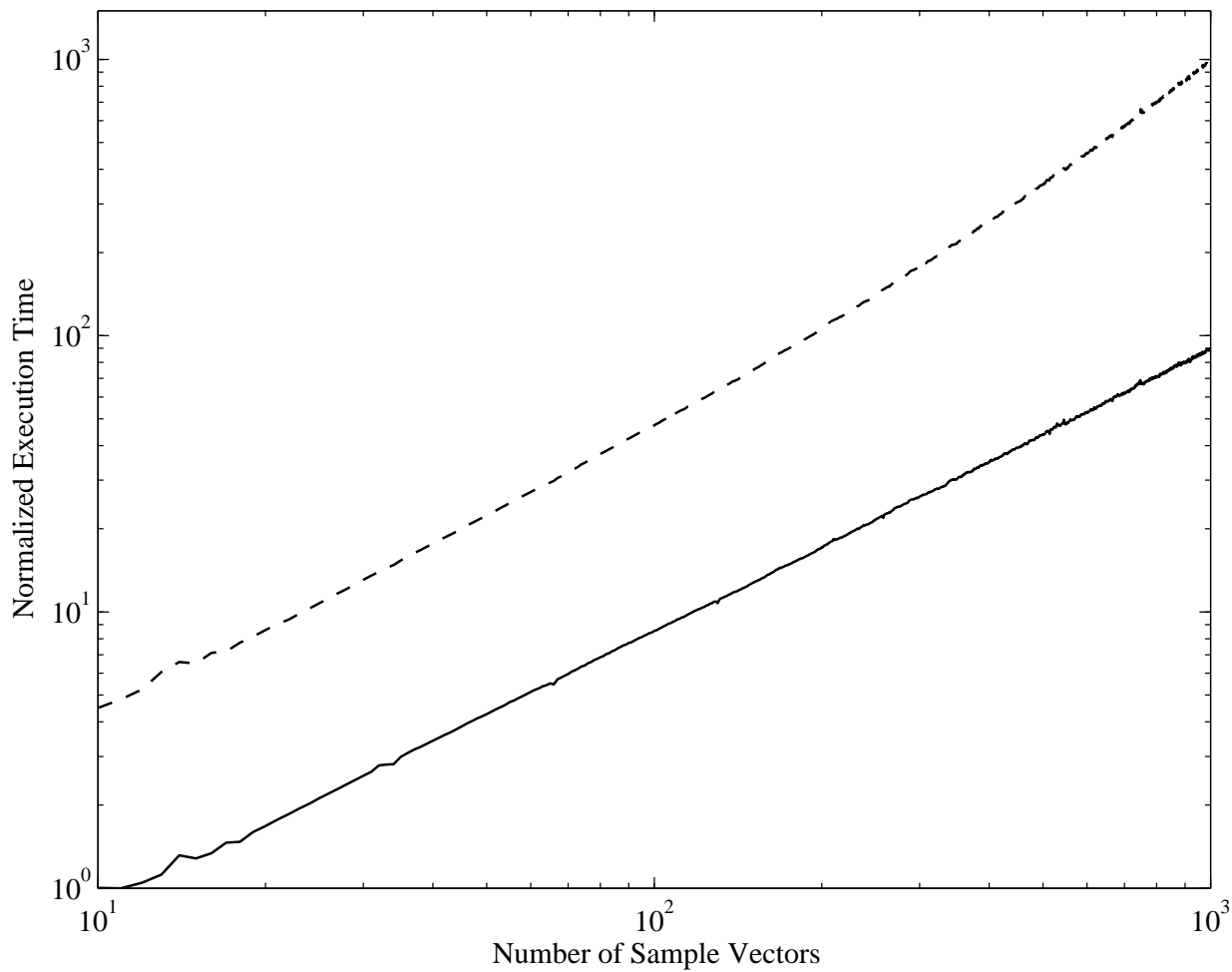}
	\end{center}
	\caption{Normalized execution time of the calculation of the covariance matrices of all subsets of $N-2$, 10 pixel, sample vectors from a set of $N$ vectors, averaged over 100 executions.  The dashed curve represents the (geometrically increasing) execution time of code which calculates each covariance from scratch, whereas the solid curve represents the (linearly increasing) execution time of code which uses update equations (\ref{eq:covplus}) and (\ref{eq:covminus}). \label{fig:exectime}}
\end{figure}

\begin{figure}
	\begin{center}
	\includegraphics[width=1\textwidth]{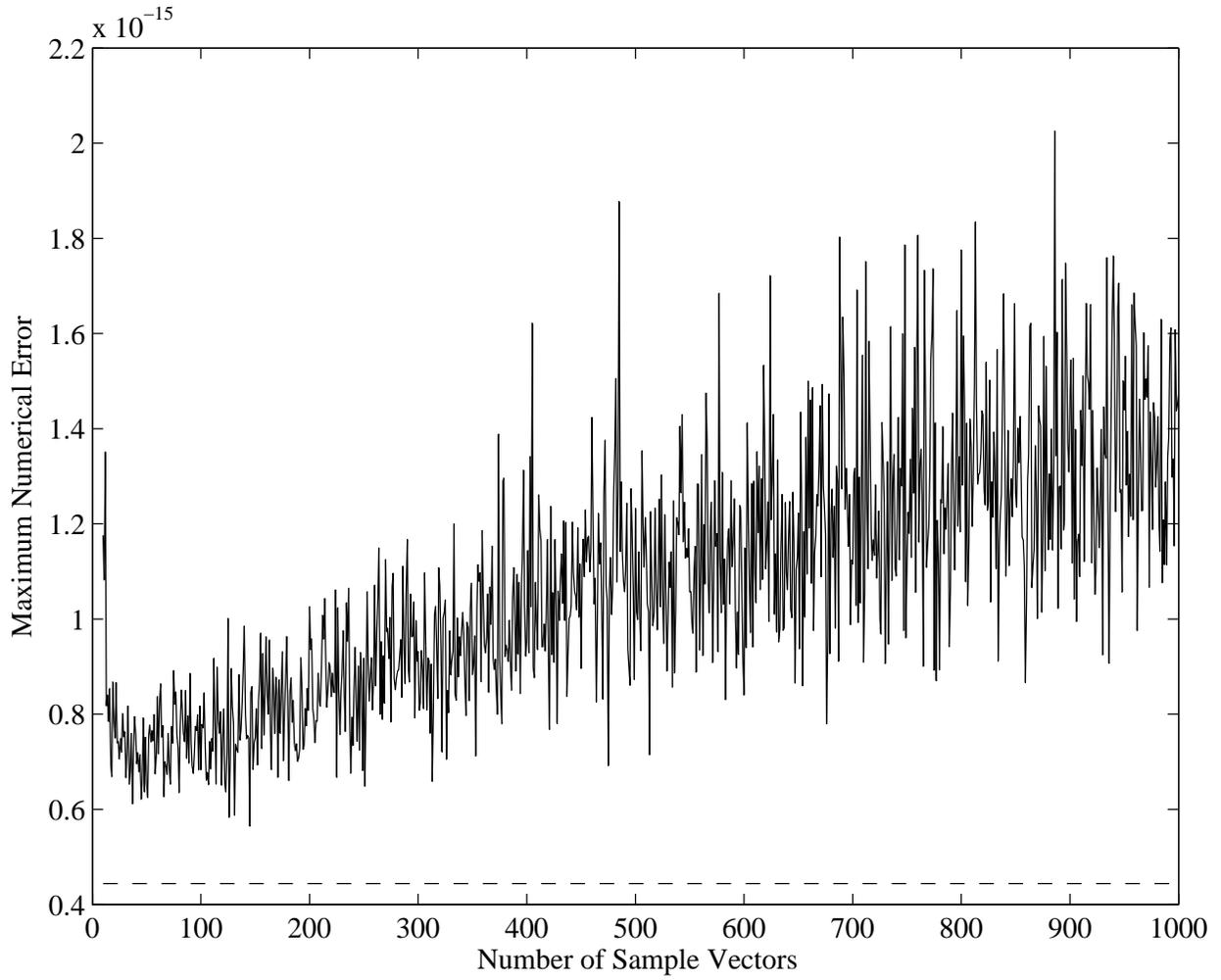}
	\end{center}
	\caption{Maximum difference between covariance values calculated by the two programs in Fig.~\ref{fig:exectime}.  The dashed line represents the precision of the data type used. \label{fig:err}}
\end{figure}

Figure \ref{fig:exectime} demonstrates the utility of these equations by comparing the execution time of two programs (running in the same environment), which calculate all of the covariances of subsets of sample vectors drawn from an increasing data set.  Note that both axes of this figure use logarithmic scales. The first program, represented by the dashed curve, calculates all covariances from scratch, and has geometrically increasing execution time.  The second program, represented by the solid curve, uses equations (\ref{eq:covplus}) and (\ref{eq:covminus}) to update covariances and has strictly linearly increasing execution time.  All times are normalized by the minimum execution time of the second program.  It is clear that in applications where we must continuously evaluate covariances of closely related data sets, this approach can lead to significant decreases in computation time.  Figure \ref{fig:err} plots the maximum differences between covariances calculated by the two codes, which increase with the number of sample vectors, but remain within a factor of 10 of the data type precision.

\section{Conclusions}
We have presented a group of techniques that can significantly improve the efficiency of calculations associated with some of the most frequently employed post-processing techniques for planetary imaging.  In particular, the introduction of sequential and neighboring covariance calculations can turn highly intensive calculations into relatively simple processes that can be run on conventional hardware in real time.  The increased computational speed has the additional benefit of allowing the user to more freely vary other parameters in the computation, which can be very important for algorithms such as LOCI where there are many additional inputs in the form of the selected subtraction and optimization zones. 

\acknowledgements
Many thanks to the attendees of the 2012 Exoplanet Imaging Workshop, and especially Lisa Poyneer, Harry Barret, Luca Caucci, Laurent Pueyo, Remi Soummer, Christian Marois and Dimitri Mawet for many useful discussions and patient explanations.  Thanks also to Bruce Macintosh for his very helpful comments and suggestions and to Jean-Baptiste Ruffio for his feedback. Portions of this work were performed under the auspices of the U.S. Department of Energy by Lawrence Livermore National Laboratory under Contract DE-AC52-07NA27344.

\bibliography{Main} 
\bibliographystyle{apj}

\end{document}